\documentclass[10pt,prc,aps,nofootinbib,showkeys,showpacs,twocolumn,superscriptaddress]{revtex4-1}
\usepackage{graphicx}
\usepackage{epsfig}
\usepackage[dvipdfm,colorlinks=true,urlcolor=blue,citecolor=blue,breaklinks=true]{hyperref}

\begin{document}
\title{On the formulation of functional theory for pairing with particle number restoration}

\author{Guillaume Hupin} \email{hupin@ganil.fr}
\affiliation{Grand Acc\'el\'erateur National d'Ions Lourds (GANIL), CEA/DSM-CNRS/IN2P3, Bvd Henri Becquerel, 14076 Caen, France}
\author{Denis Lacroix} \email{lacroix@ganil.fr}
\affiliation{Grand Acc\'el\'erateur National d'Ions Lourds (GANIL), CEA/DSM-CNRS/IN2P3, Bvd Henri Becquerel, 14076 Caen, France}
\author{Michael Bender}
\email{bender@cenbg.in2p3.fr}
\affiliation{Universit{\'e} Bordeaux,
             Centre d'Etudes Nucl{\'e}aires de Bordeaux Gradignan, UMR5797,
             F-33175 Gradignan, France}
\affiliation{CNRS/IN2P3,
             Centre d'Etudes Nucl{\'e}aires de Bordeaux Gradignan, UMR5797,
             F-33175 Gradignan, France}

\begin{abstract}
The restoration of particle number within Energy Density Functional theory is 
analyzed. It is shown that the standard method based on configuration mixing
leads to a functional of both the projected and non-projected densities. As an alternative 
that might be advantageous for mass models, nuclear dynamics and thermodynamics, we propose 
to formulate the functional in terms directly of the one-body and two-body density matrices of the 
state with good particle number.
Our approach does not contain the pathologies recently observed 
when restoring the particle number in an Energy Density Functional framework based on
transition density matrices and can eventually be applied with functionals 
having arbitrary density dependencies.        
\end{abstract}

\date{28 April 2011}


\maketitle

\section{Introduction}
\label{sec:intro}

Energy Density Functional (EDF) methods provide a universal framework to  
describe nuclear structure, nuclear dynamics or thermodynamics. 
Tremendous advances have been made in the last few decades on the 
practical application of EDF methods to nuclei \cite{Ben03}. 
Still, despite their long success, some of the fundamental assumptions 
made to justify the usual strategies how the EDF techniques are constructed 
and used for nuclear systems have not yet been satisfactorily clarified. 
Most, if not all, EDF approaches break as many symmetries of the nuclear 
Hamiltonian as possible: translational, rotational and $U(1)$ symmetry 
in gauge space, among the most important ones.  In fact, the exploitation 
of symmetry breaking in nuclei is strongly motivated by experimental 
observations. For instance, the appearance of highly collective rotational 
bands in spectroscopic data clearly points to the existence of deformed 
intrinsic states in many nuclei \cite{Rin80}. Similarly, there is evidence 
that pairing can be often treated by explicitly breaking the $U(1)$ gauge 
symmetry of eigenstates of the particle-number operator, like for instance 
in a Bardeen-Cooper-Schrieffer (BCS) or Hartree-Fock-Bogoliubov (HFB) 
approach \cite{Bla86,Rin80}. Nuclei are, however, finite systems and methods 
like BCS or HFB do not properly treat quantum fluctuations of the order 
parameter associated with the broken symmetry \cite{Bla86}. These fluctuations 
can be incorporated either by a statistical treatment of the order parameter, 
or by the restoration of the relevant symmetry  \cite{Bla86}. 
The concept of symmetry breaking and restoration stands out as the tool of 
choice within the EDF framework. 

It has, however, been recently shown that restoration of symmetries has 
to be handled with great care in an EDF framework \cite{Ang01,Dob07,Lac09a,Dug10}.
In particular, the configuration mixing within a Multi-Reference (MR) EDF approach
might lead to serious practical difficulties that can, however, eventually 
be cured \cite{Lac09a,Ben09}. Besides compromising applications, 
these difficulties have clearly pointed out the necessity to clarify the
theoretical framework on which the theory can be build.

The discussion in the present paper is restricted to ground-state properties
and to particle-number projection, for which detailed analyses have 
been recently made. This case is the simplest situation where pathologies of 
the MR-EDF approach have been observed \cite{Ang01}, analyzed and 
regularized \cite{Lac09a,Ben09}. The first goal of the present work is 
to provide an alternative analysis of the EDF theory using configuration 
mixing to restore symmetries without and with the regularization.
It will be shown that neither the non-regularized nor the regularized 
functionals can straightforwardly be interpreted in terms of the densities 
of projected or non-projected states. Starting from this analysis, the 
second intent of this work is to propose an alternative way to introduce 
a functional theory that is $U(1)$ symmetry conserving, and that without 
making use of the Multi-Reference technique. Our approach remains close 
to the Hohenberg-Kohn \cite{Hoh64} and Kohn-Sham \cite{Koh65}
framework and uses a projected state as an intermediate trial state to 
construct the components of the functional. This approach avoids the 
difficulties recently encountered in MR-EDF approaches and can be applied 
also with functionals that cannot be safely employed within the standard 
MR-EDF approach, as for example functionals with non-analytical density 
dependences.

\section{Particle number restoration within EDF theory: standard approach}
\label{sec:edfstandard}

The strategy to obtain a functional for pairing including particle-number 
restoration has been extensively analyzed recently \cite{Dob07,Lac09a,Ben09,Dug09} 
and we only give here the main steps necessary for our discussion.  
Following these references, in this section we will consider a specific 
class of functional form that will be sufficient for the present 
discussion.\footnote{
Note that, none of the currently used SR-EDF functionals 
belongs to this class as they have non-analytical density dependences. The 
form~(\ref{eq:denssr}) is the only one (restricting ourselves here to 
bilinear functionals) for which the  recently proposed regularization applies.
}
At the so-called Single-Reference (SR) level, a quasi-particle (QP) vacuum of 
Bogoliubov type $| \Phi_0 \rangle$ is used to construct the normal and 
anomalous density matrices, denoted by $\rho$ and $\kappa$, that serve 
to construct the functional. The energy is then written as   
\begin{eqnarray}
{\cal E}_{SR}[\Phi_0] =&& \mathcal{E}_{SR} \left[ \rho , \kappa, \kappa^* \right] \nonumber \\
=&&  \sum_{i} t_{ii} \rho_{ii}+ \frac{1}{2} \sum_{i,j} \overline{v}_{ijij}^{\rho \rho}  \rho_{ii}\rho_{jj}  \nonumber \\
&+& \frac{1}{4} 
 \sum_{i,j} \overline{v}_{i\bar\imath j\bar\jmath }^{\kappa \kappa} \kappa_{i \bar\imath }^* \kappa_{j \bar\jmath } \, ,
\label{eq:denssr}
\end{eqnarray}
where $\overline{v}^{\rho \rho}$ and $\overline{v}^{\kappa \kappa}$ denote the effective vertices 
in the particle-hole and particle-particle channels. Standard SR-EDF can be schematically seen as
the sequence
\begin{equation}
\label{eq:seqSR}
\Phi_0 ~\Longrightarrow~ ( \rho , \kappa, \kappa^*) ~\Longrightarrow ~ {\cal E}_{SR}  \, .
\end{equation}
The price to be paid for incorporating pairing with a rather simple functional is to use an 
intermediate state $| \Phi_0 \rangle$ that is not an eigenstate 
of particle number. In a second step, the symmetry can be restored projecting out 
the component with $N$ particles
\begin{eqnarray}
| \Psi_N \rangle = P^N | \Phi_0 \rangle \, ,
\end{eqnarray}
where $P^N$ denotes the particle number projection operator defined through \cite{Rin80,Bla86}
\begin{equation}
\label{eq:Pop}
{P}^N
=  \frac{1}{2\pi} \int_{0}^{2\pi} \! d{\varphi} \; \,e^{i\varphi (\hat{N}-N)}
\, .
\end{equation}
The expectation value of any operator $O$ that conserves particle number can then be 
expressed as 
\begin{equation}
\label{eq:expproj}
\frac{\langle \Psi_{N} | \, O \, | \Psi_{N} \rangle}{\langle \Psi_{N} |  \Psi_{N} \rangle}
=  \int_{0}^{2\pi}  d\varphi \, 
 \frac{ \langle \Phi _0 | O  | \Phi_\varphi \rangle } {\langle \Phi _0 | \Phi_\varphi \rangle}  {\cal N}_N ({0, \varphi}) 
\, ,
\end{equation} 
where the shorthand  
\begin{eqnarray}
{\cal N}_N ({0, \varphi}) \equiv  \frac{e^{-i\varphi N}}{2\pi}
 \frac{\langle  \Phi_0 | \Phi _{\varphi} \rangle}{\langle \Psi_{N} |  \Psi_{N} \rangle}\, ,
\end{eqnarray}
has been introduced. Here $\varphi$ denotes the gauge angle, whereas 
$| \Phi_{\varphi} \rangle = e^{i \varphi \hat N} \, | \Phi_0 \rangle$ 
refers to the state  $| \Phi_0 \rangle$ rotated in gauge space. The 
kernel entering in the integral of Eq.~(\ref{eq:expproj}) corresponds 
to the transition matrix element of an operator between two quasi-particle 
vacua. One can then take advantage of  the Generalized Wick Theorem (GWT) 
\cite{Bal69,Rin80} to express the kernel in terms 
of the transition density matrices
\begin{eqnarray}
\label{contractph}
\rho^{0\varphi}_{i j }
& \equiv & \frac{\langle \Phi_{0} | a^{\dagger}_{j} a_{i} | \Phi_{\varphi} \rangle}
           {\langle \Phi_{0} | \Phi_{\varphi} \rangle} ,     \\
\label{contracthh}
\kappa^{0\varphi}_{i j }
& \equiv & \frac{\langle \Phi_{0} | a_{j} a_{i} | \Phi_{\varphi} \rangle}
           {\langle \Phi_{0} | \Phi_{\varphi} \rangle},
      \\
\label{contractpp}
{\kappa^{\varphi 0 }_{j i}}^\star
& \equiv & \frac{\langle \Phi_{0} | a^{\dagger}_{i } a^{\dagger}_{j} | \Phi_{\varphi} \rangle}
           {\langle \Phi_{0} | \Phi_{\varphi} \rangle}.
\end{eqnarray}
For instance, when $O$ is a two-body Hamiltonian, the two-body interaction 
$\overline{v}$ entering in (\ref{eq:expproj}) takes a form similar 
to Eq.~(\ref{eq:denssr}) with $\overline{v}^{\rho\rho} = 
\overline{v}^{\kappa \kappa}= \overline{v}$ and where the densities, $\rho$ and $\kappa$ 
are replaced by the corresponding transition densities, Eqns.~(\ref{contractph}-\ref{contractpp}). 
Guided by the Hamiltonian case, the energy functional associated with particle-number 
restoration is usually defined through
\begin{eqnarray}
\label{eq:ekernel}
{\cal E}_N [\Psi_{N}] \equiv   \int_{0}^{2\pi}  d\varphi \, \mathcal{E}_{SR} \left[ \rho^{0\varphi} , \kappa^{0\varphi}, 
{\kappa^{\varphi 0 }}^\star \right]
   {\cal N}_N ({0, \varphi}) \, .  
\end{eqnarray}
This energy functional is a special case of a so-called Multi-Reference 
EDF (MR-EDF). The present strategy to restore symmetries in an EDF 
framework deserves some further remarks:
\begin{itemize}
\item  
  The expression (\ref{eq:ekernel}) is postulated having in mind the Hamiltonian case. 
  However, the MR-EDF theory should not 
  be confounded with the expectation value of a Hamilton operator. In particular, an 
  energy functional has much more flexibility regarding the functional form of the 
  energy kernels in Eq.~(\ref{eq:ekernel}), which can be used for the efficient
  modeling of in-medium correlations.
\item  
  The construction of the MR-EDF, Eq.~(\ref{eq:ekernel}), from the SR-EDF by 
  simply replacing the normal and anomalous density matrices in the SR EDF 
  by the corresponding transition density matrices is postulated by analogy to 
  the GWT. While it appears rather natural, it was shown recently that this strategy 
  to construct the MR-EDF might lead to an ill-defined functional that exhibits
  divergencies and jumps in practical applications~\cite{Dob07,Lac09a,Ben09}. 
  While a solution to this problem has been proposed and applied in Refs.~\cite{Lac09a,Ben09}, 
  a consistent framework for MR-EDF approaches is still missing. 
  A clear illustration of this is the ongoing debate about  which densities should enter 
  in the functional~\cite{Rob10}, as well as the recently recognized impossibility to 
  use non-integer powers of the transition density in MR energy functionals \cite{Dug09}.
  \item The very notion of symmetry restoration within an EDF framework remains to be clarified. 
  For instance, it has been shown recently~\cite{Dug10} that also regularized MR energy 
  functionals may in general not transform as an irreducible representation of the restored 
  symmetry, unless additional constraints are introduced.
\end{itemize}
In the present section, we will further analyze the way the MR-EDF is constructed 
and the possible sources of difficulties. For simplicity, we restrict ourselves to 
a case where the two-body effective interaction  kernels entering Eq.~(\ref{eq:denssr}) 
are {\it independent of the densities}. 

A peculiarity of particle-number projection is that the canonical basis of the original 
state $| \Phi_0 \rangle$ and of the rotated states $| \Phi_\varphi \rangle$ are the 
same when making a suitable choice of the Bogoliubov transformation between
quasi-particle operators of these states. Accordingly, the canonical base of the 
projected state $| \Psi_N \rangle$ is also the same as the one of the original 
reference state $| \Phi_0 \rangle$. In the following, we will implicitly assume 
that densities are written in this canonical basis in which we have
\begin{equation}
\label{eq:canodens}
\rho^{0\varphi}_{i j } = \delta_{ij} n^{0\varphi}_{i}, ~~~\kappa^{0\varphi}_{i j } = \delta_{j \bar\imath } \kappa^{0\varphi}_{i \bar\imath }, 
~~~{\kappa^{\varphi 0 }_{i j }}^\star = \delta_{j \bar\imath } {\kappa^{\varphi 0 }_{i \bar\imath }}^\star \,, 
\end{equation}  
whereas the energy $ {\cal E}_{N} $ takes the form
\begin{eqnarray}
 {\cal E}_{N}  [\Psi_{N}] &=& \sum_{i} t_{ii}  \int_{0}^{2\pi} \!\!\! d\varphi  ~n^{0\varphi}_i  {\cal N}_N(0,\varphi)
 \nonumber \\
 &+& \frac{1}{2} \sum_{i,j} \overline{v}_{ijij}^{\rho \rho}    \int_{0}^{2\pi} \!\!\! d\varphi  ~n^{0\varphi}_i ~n^{0\varphi}_j   {\cal N}_N({0,\varphi}) \nonumber \\
 &+&\frac{1}{4} \sum_{i, j} \overline{v}_{i\bar\imath j\bar\jmath }^{\kappa \kappa}  
\int_{0}^{2\pi} \!\!\! d\varphi ~{\kappa^{\varphi 0}_{i \bar\imath }}^\star {\kappa^{0\varphi}_{j \bar\jmath }}   \, {\cal N}_N({0,\varphi}) \, .
\end{eqnarray}  
After a lengthy, but straightforward calculation, the energy functional can be expressed as
\begin{eqnarray}
 {\cal E}_{N} [\Psi_{N}] &=& \sum_{i} t_{ii} n^N_i \nonumber \\
 &+& \frac{1}{2} \sum_{i,j, j\neq \bar\imath } \overline{v}_{ijij}^{\rho \rho}   R^N_{ijij} \nonumber \\
 &+& \frac{1}{4} 
\sum_{i \neq j, i  \neq \bar \jmath } \overline{v}_{i\bar\imath j\bar\jmath }^{\kappa \kappa}  R^N_{j \bar\jmath  i\bar\imath  }  \nonumber \\
&+& \frac{1}{2} \sum_{i} \overline{v}_{i\bar\imath  i \bar\imath }^{\rho \rho}  
 \int_{0}^{2\pi} \!\!\! d\varphi  ~n^{0\varphi}_i ~n^{0\varphi}_i   {\cal N}_N({0,\varphi}) \nonumber \\
 &+&\frac{1}{2} \sum_{i}\overline{v}_{i\bar\imath i\bar\imath }^{\kappa \kappa} 
 \int_{0}^{2\pi} \!\!\! d\varphi {\kappa^{\varphi 0}_{i \bar\imath }}^\star {\kappa^{0\varphi}_{i \bar\imath }}   \, {\cal N}_N({0,\varphi}) \, ,
 \label{eq:edftotproj}
\end{eqnarray} 
where $n^N_i$ are the occupation numbers:
\begin{eqnarray}
n^N_i &\equiv& 
\frac{ \langle \Psi_{N} | a^\dagger_i a_i | \Psi_{N} \rangle }{\langle \Psi_{N} | \Psi_{N} \rangle} 
\end{eqnarray}
and $R^N_{ijkl}$ corresponds to the two-body density matrix 
\begin{eqnarray}
R^N_{ijkl} &\equiv& \frac{ \langle \Psi_{N} | a^\dagger_k a^\dagger_l a_j a_i | \Psi_{N} \rangle }{\langle \Psi_{N} | \Psi_{N} \rangle} 
\end{eqnarray} 
of the projected state. They can be expressed in terms of the gauge angle integrals as
\begin{eqnarray}
n^N_i 
&=&  \int_{0}^{2\pi}  d\varphi  \,  ~ n^{0\varphi}_i {\cal N}_N({0,\varphi}) \, ,
\end{eqnarray}
and
\begin{eqnarray}
R^N_{ijkl} 
&=&  (\delta_{ik} \delta_{jl} -  \delta_{il} \delta_{jk}) \int_{0}^{2\pi} \!\!\! d\varphi ~ n^{0\varphi}_i n^{0\varphi}_j   {\cal N}_N({0,\varphi})  \nonumber \\
&+&  \delta_{j \bar\imath } \delta_{l \bar k}  \int_{0}^{2\pi} \!\!\! d\varphi {\kappa^{\varphi 0}_{i \bar\imath }}^\star {\kappa^{0\varphi}_{k \bar k}}    \, {\cal N}_N({0,\varphi}) \, .
\end{eqnarray} 
Equation~(\ref{eq:edftotproj}) is rather enlightening with respect to the 
physical content of present MR-EDF calculations.
Indeed, if one neglects the last two terms in Eq. (\ref{eq:edftotproj}), one sees that the functional associated with the projected state 
can be regarded as a functional of the one- and two-body components of this very state. Similarly, 
if one uses the same effective interaction $\overline{v}^{\rho\rho} = 
\overline{v}^{\kappa \kappa}$, then the last two terms of Eq. (\ref{eq:edftotproj}) recombine and the two-body component  $ R^N_{i\bar\imath  i \bar\imath  }$ can be 
recognized, thanks to the relation
\begin{eqnarray}
  R^N_{i\bar\imath  i \bar\imath  } &=&  n_i^N \nonumber \\
  &=& \int_{0}^{2\pi} \!\!\! d\varphi  ~(n^{0\varphi}_i ~n^{0\varphi}_i  + {\kappa^{\varphi 0}_{i \bar\imath }}^\star {\kappa^{0\varphi}_{i \bar\imath }}) {\cal N}_N({0,\varphi}) \, . \label{eq:sumrule}
\end{eqnarray}
However, when using different effective vertices $\overline{v}^{\rho\rho} \neq
\overline{v}^{\kappa \kappa}$ in the particle-hole and particle-particle channels, or when using 
vertices $\overline{v}^{\rho\rho}$ or $\overline{v}^{\kappa \kappa}$ that cannot be written as an 
antisymmetrized matrix elements of the two-body force, then the identification of the energy as 
a functional of one- and two-body density matrices of the projected state cannot be made 
anymore. Instead, it can only be written as a functional of the transition density 
matrices.\footnote{We recall that the expectation value of the two-body operator in a 
projected state can be written as a functional of the two-body density of this state, 
or, fully equivalently, as a functional of the one-body density matrices. This property 
does not hold for general functionals  that are constructed without reference to an 
underlying Hamiltonian.
}
This subtlety is intimately connected to the presence of pathologies encountered 
in MR-EDF calculations.  Indeed, the last two terms in Eq.~(\ref{eq:edftotproj})
are nothing but the ones at the heart of the difficulties to construct a well-defined 
MR-EDF theory. As discussed in Refs.~\cite{Dob07,Ben09}, for near-orthogonal states 
$\langle \Phi _0 | \Phi_\varphi \rangle  \simeq 0$ there is at least 
one $n_i^{0\varphi}$ and the corresponding ${\kappa^{\varphi 0}_{i \bar\imath }}^\star$ 
and ${\kappa^{0\varphi}_{i \bar\imath }}$ that all go to infinity. As a consequence, 
the two terms  can separately become larger than any physical scale in the nucleus. 
They do, however, recombine to a well-behaved expression  when a Hamiltonian is used, 
i.e.\ when $\overline{v}^{\rho\rho} = \overline{v}^{\kappa \kappa}$. Without taking 
specific care of these terms in the restoration of symmetry within the functional 
framework, there is a spurious contribution that leads to discontinuities and 
divergences when plotting the particle-number projected energy as a function of 
a collective coordinate.

\subsection{MR-EDF theory with regularization}

A strategy to construct a well-behaved MR-EDF theory proposed in Refs.~\cite{Lac09a,Ben09} 
is to remove terms that might  not properly recombine in the MR-EDF approach in 
such a way that the spurious contamination is removed 
without touching the physical content of the functional. 
The resulting functional then takes the form (technical details are given in 
appendix~\ref{app:correctedmr})
\begin{eqnarray}
 {\cal E}_{N} [\Psi_{N}] &=& \sum_{i} t_{ii} n^N_i \nonumber \\
&+& \frac{1}{2} \sum_{i,j, j\neq \bar\imath } \overline{v}_{ijij}^{\rho \rho}   R^N_{ijij} \nonumber \\
&+& \frac{1}{4} 
\sum_{i \neq j, j\neq \bar\imath } \overline{v}_{i\bar\imath j\bar\jmath }^{\kappa \kappa}  R^N_{j \bar\jmath  i\bar\imath  }  \nonumber \\
&+& \frac{1}{2} \sum_{i} \overline{v}_{i\bar\imath  i \bar\imath }^{\rho \rho}  (n^N_i n^N_i - \delta n_i \delta n_i) \nonumber \\
 &+&\frac{1}{2} \sum_{i}\overline{v}_{i\bar\imath i\bar\imath }^{\kappa \kappa} \left[ n^N_i (1 - n^N_i) + \delta n_i \delta n_i \right]  \,  ,
 \label{eq:edftotprojcor}
\end{eqnarray}  
where $\delta n_i = n^N_i -n^0_i$ is the difference between the occupation number 
of the level $i$ in the projected and the non-projected state. 

Expression (\ref{eq:edftotprojcor}) is of particular interest for the following discussion 
regarding the construction 
of energy functional theory. First, let us remark that, compared to the previous form 
 (\ref{eq:edftotproj}), the gauge space integrals are now hidden in the components of the one- 
and two-body density matrices of the projected state. In addition, the last two lines 
of Eq.~(\ref{eq:edftotprojcor}) are also functionals of the  occupation numbers 
$n^0_i$ in the original non-projected state.  
The analysis of the regularization procedure to remove spurious contribution 
to the MR-EDF method~\cite{Lac09a,Ben09,Dug09} suggests  that these terms will always be well-behaved.

\begin{figure}[t!]
\includegraphics[width=8.cm]{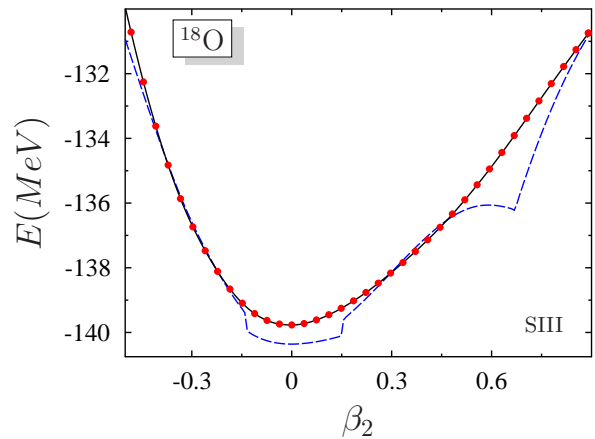}
\caption{\label{fig1:pav}
(Color online)
Particle-number restored deformation energy curve of $^{18}$O as a function of quadrupole 
deformation $\beta_2$ and calculated 
within standard MR-EDF technique using Projection 
After Variation (PAV) with
SIII and a delta pairing interaction before (blue dashed curve) and after regularization (black solid curve).
The red filled circles correspond to the result 
obtained using directly Eq.~(\ref{eq:edftotprojcor}) (see text).
To compare with previous work \cite{Ben09}, the Coulomb exchange contribution has been subtracted from the energy.
}
\end{figure}

An example for a deformation energy curve obtained from a particle-number projected 
MR-EDF calculation with the Skyrme interaction SIII and a pairing functional of volume type 
is shown in Fig.~\ref{fig1:pav} (dashed line). The MR-EDF is numerically calculated
using expression~(\ref{eq:ekernel}) and the Fomenko discretization procedure of
the gauge-space integrals described, for instance, in Ref.~\cite{Ben09}. Here, 199 
discretization points have been used. This large number is necessary to resolve the 
discontinuities that stem from the spurious contribution to the non-regularized 
MR-EDF \cite{Ben09}.
As in Ref.~\cite{Ben09}, the Lipkin-Nogami procedure is used in the minimization 
of the energy of the state $| \Phi_0 \rangle$. The solid line corresponds to the 
MR-EDF method with the regularization proposed in~\cite{Lac09a}. In this Figure, 
we also show the results (filled circles) obtained using directly the expression~(\ref{eq:edftotprojcor}) 
that has been proven above to be analytically equivalent to the regularized MR-EDF functional. Note that, 
in the latter case, we have used a method called hereafter "recurrence 
method" to compute the projected occupation numbers and components of the projected two-body densities. 
This method is described in detail in appendix~\ref{app:rec}. 
Although  the use of gauge angle integration would have given exactly the same results, this method has the advantage 
to be very simple, numerically efficient and to not make use of transition density matrices. 
As expected, the energy obtained with expression~(\ref{eq:edftotprojcor}) exactly 
matches the one obtained using the regularized MR-EDF functional. 
This formulation 
provides a new and alternative  insight  into the content of particle-number
restored energy functionals.

\subsection{Critical analyses of standard method}

As discussed above, specific regularizations in MR-EDF functionals are needed to 
avoid discontinuities such as the jumps appearing in Fig.~\ref{fig1:pav}. At this point, 
even with the regularization, two important problems remain: 
\begin{itemize}
  \item[(i)] Terms that have a non-analytical density dependence, for example a 
  non-integer power such as often used in parameterizations of the Skyrme and Gogny 
  interactions, cannot be regularized with the procedure proposed in 
  Ref.~\cite{Lac09a}. Indeed, the functional itself becomes in that case multivalued 
  in the complex plane and cannot be properly defined~\cite{Dob07,Dug09}. 
  \item[(ii)] A second issue illustrated in Eq.~(\ref{eq:edftotprojcor}) is that the last 
  two terms are not only a functional of the occupation numbers of the projected state, 
  but also of the occupation probabilities of the original reference state 
  $| \Phi_0 \rangle$. Accordingly, the energy remains a functional of the density 
  of the quasi-particle vacuum that is not an eigenstate of particle number. 
  This raises the question which density, i.e.\ projected, transition, or 
  non-projected can be used to construct a functional for MR calculations.
\end{itemize}

In the following, we show that both (i) and (ii) can eventually be avoided by changing 
the strategy to construct the functional for pairing that accounts for particle number 
restoration.

\section{Discussion on EDF theory for pairing with particle number restoration}

Let us now discuss the critique (ii) made above concerning the 
components of the projected energy functional. 
In most functional approaches, an intermediate 
state is introduced to construct densities that are used to minimize the energy.
This is the case in usual 
DFT or at the SR-EDF level where the trial state is a Slater determinant or 
a quasi-particle vacuum. When restoring the symmetry in a MR-EDF framework, then,
according to Eq.~(\ref{eq:edftotprojcor}), the projected state can  be {\it almost} 
regarded as an intermediate many-body state from which the one- and two-body density 
matrices used to define the functional are obtained.

However, due to the presence of $n^0_i$ in the energy, this functional happens 
to depend on components not only of the projected state, but also of the original 
reference state. A slight  modification, however, can easily restore the unique 
dependence of the functional on the projected state.  
If, for instance, the following replacements
\begin{eqnarray}
\label{eq:nidnini}
\begin{array}{lcl}
(n^N_i n^N_i - \delta n_i \delta n_i)       &   \Longrightarrow & n^N_i n^N_i  \\
\\
\left[ n^N_i (1 - n^N_i) + \delta n_i \delta n_i \right]      &    \Longrightarrow & n^N_i (1 - n^N_i) \, ,
\end{array}
\end{eqnarray}
are made in Eq.~(\ref{eq:edftotprojcor}), then the strategy of standard 
DFT to construct the EDF as a functional  of a density of an auxiliary state, 
the projected state here, is recovered.\footnote{This does not mean, however,
 that we recover a theory that is equivalent to DFT. Indeed, at this stage, 
the functional~(\ref{eq:edftotprojcor}) is  still a functional of the two-body 
density matrix. However, as will be discussed below, for the specific case 
 of particle-number projection, the two-body density matrix is itself a 
functional of the one-body density matrix.}  

\begin{figure}[t!]
\includegraphics[width=8.cm]{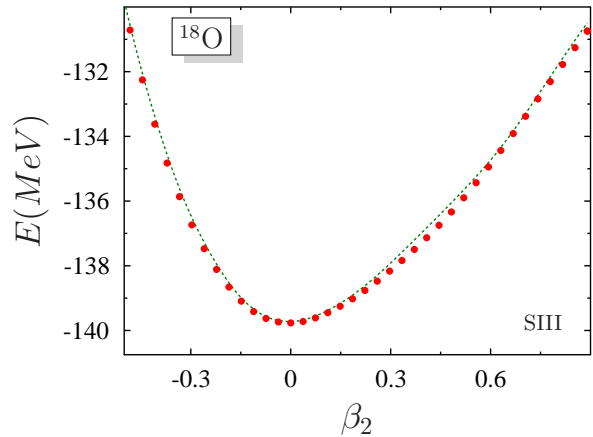}
\caption{\label{fig2:pav}
(Color online)
Particle-number restored deformation energy surface 
of $^{18}$O calculated using Eq.~(\ref{eq:edftotprojcor}). 
The dotted curve is obtained by making the replacement~(\ref{eq:nidnini}) 
in Eq.~(\ref{eq:edftotprojcor}).
}
\end{figure}

The use of a projected product state a auxiliary state has the advantage that it allows to 
treat pairing in a particle-number conserving framework.
An illustration of a result obtained taking into account this modification in Eq.~(\ref{eq:edftotprojcor})   
is shown in Fig.~\ref{fig2:pav} and compared to the original curve. This figure illustrates that 
the small change in the functional does not significantly modify the energy landscape. This is
indeed not unexpected since the difference $\delta n_i$ (resp.\ $\delta n_i \delta n_i$) is likely to be much
smaller than $n^N_i$ (resp.\ $n^N_i n^N_i$).

By making the simple modification~(\ref{eq:nidnini}), the EDF framework can now 
be interpreted as a functional of the projected-state densities. Indeed, 
the state with good particle number can now be regarded as the auxiliary many-body 
state that provides the quantities used to construct the EDF. Similarly 
to Eq.~(\ref{eq:seqSR}), the corresponding theory can be regarded as a new sequence
\begin{equation}
\label{eq:seqN}
\Psi_N~\Longrightarrow~ ( \rho^N , R^N) ~\Longrightarrow ~ {\cal E}_{N}.
\end{equation}
We note in passing that the slight modification~(\ref{eq:nidnini}) does not break the shift invariance 
of the energy functional  discussed in Refs.~\cite{Dob07,Ben09}.
{
At this point, let us make a few further important remarks:
\begin{itemize}
\item The functional form~(\ref{eq:edftotprojcor}) is not completely surprising. 
Indeed, our starting point, Eq.~(\ref{eq:denssr}), is very close to a form one would 
have obtained by taking the expectation value of a two-body Hamiltonian. In the 
case of an energy functional calculated 
as the expectation value of a genuine Hamiltonian operator, the energy can be written 
as a functional of one-body transition densities, but also as a functional of projected 
one- and two-body (and eventually higher-order) densities. This freedom is lost in the 
functional framework, where a choice has to be made for either one or the other. MR-EDF 
follows the former strategy, whereas the latter has not been explored yet. For a 
regularized bilinear functional, the differences 
between both formulations remain very small, see Fig.~\ref{fig2:pav}.   
  
\item  { 
Expression~(\ref{eq:edftotprojcor}) contains not only one-body but also two-body 
matrix elements and might appear out of the scope of a density functional approach 
aiming at replacing the original $N$-body problem by a functional of the one-body 
density matrix only. Indeed, in the Hamiltonian case, the expectation value of any
two-body Hamiltonian for any state can directly be regarded as a functional of the 
one- and two-body densities of this state. Density functional theories are introduced 
to avoid the explicit use of two-body and higher density matrices. Therefore, by 
itself, the use of a functional of the two-body density might appear useless. The 
important simplification here stems from the fact that these densities are constructed 
from a very specific class of states, namely projected product states. For instance, 
we have shown recently that the two-body density matrix elements
can eventually be written as an explicit functional of the one-body density  
under some approximation~\cite{Lac10}. Accordingly, while two-body density matrix elements are used 
to get a compact expression in Eq.~(\ref{eq:edftotprojcor}), this functional can truly 
be considered as a functional of the projected state one-body density consistently 
with density matrix functional theory, such that the sequence become
\begin{equation}
\label{eq:seqNone}
\Psi_N~\Longrightarrow~ \rho^N  ~\Longrightarrow ~ {\cal E}_{N} \, .
\end{equation} 
}
  \item When making the replacement (20) in Eq. (19), then the functional directly incorporates symmetry breaking and its restoration in a single step, contrary to standard approaches in EDF theory. From that point 
of view, it could be seen as a "Symmetry-Conserving" EDF theory.\footnote{It should 
be, however, kept in mind that the present functional only takes care of the restoration 
of $U(1)$ gauge symmetry while others still remain broken.} 
  \item It is quite interesting to note that the MR-EDF can already almost be regarded as a functional 
  of the components of the projected state.
  While this 
  was hidden in formula~(\ref{eq:ekernel}), it becomes evident in Eq.~(\ref{eq:edftotprojcor}). 
  In particular, as noted in Refs.~\cite{Lac09a,Ben09,Dug09}, there exists some 
flexibility in the regularization of the pathologies of the MR-EDF.  It is possible to slightly 
 modify the original prescription~(\ref{eq:cor_rr}-\ref{eq:cor_kk}), 
such that the regularization automatically leads to~(\ref{eq:nidnini}). In that case, the 
method based on the use of MR-EDF and "symmetry conserved" EDF framework are strictly 
equivalent. As an important consequence, while the use of techniques inspired from 
configuration mixing was unclear within a functional framework, we give here evidence
that it can be formulated consistently in a functional framework. It is, however, worth 
mentioning that while this connection can be made only in the simple functional form 
given in Eq.~(\ref{eq:denssr}), most functionals currently used do not allow their 
controlled usage in an MR EDF framework.

\item Finally, it is important to mention that this equivalence holds only true for the 
schematic bilinear functional given by Eq.~(\ref{eq:denssr}) with two-body vertices 
independent on the density. If density dependent terms are present in the functional, 
like in all currently used parameterizations of the EDF, such an equivalence does 
not exist anymore. Note, however, that, in this case, a safely usable MR-EDF cannot be
constructed anymore due to the absence of a regularization scheme. 
In Eq.~(\ref{eq:edftotprojcor}), one then obtains a functional that remains closer 
to the spirit of DFT based on the Hohenberg-Kohn theorem than the usual MR-EDF 
approach. Indeed, in the HK-theorem-based DFT, the functional is constructed 
from the density matrices of the correlated (i.e.\ in our case projected) state. 
As we will illustrate below, on the contrary, the alternative formulation proposed 
here that treats both symmetry breaking and restoration  simultaneously can still 
be applied for functionals that cannot be regularized in a MR-EDF framework.  
\end{itemize}
}

\subsection{Constraints on the symmetry-conserving functional}

If the standard projection method is used as  guidance to construct the functional, 
then the form of the functional is almost entirely constrained. Indeed, this 
corresponds to use Eq.~(\ref{eq:edftotproj}) or eventually Eq.~(\ref{eq:edftotprojcor})
as a starting point. Eq. (\ref{eq:nidnini}) corresponds to a specific choice. Here, we discuss 
whether alternative choices can be made for the last two lines of Eq.~(\ref{eq:edftotprojcor}).
At present, it is not clear if, within the functional framework,  a unique prescription of the functional
form exists. Nevertheless, one can propose a few rules to better constrain its form. 
Let us assume a more general prescription than Eq.~(\ref{eq:nidnini})
\begin{eqnarray}
\label{eq:nidninigen}
\begin{array}{lcl}
(n^N_i n^N_i - \delta n_i \delta n_i)       &   \Longrightarrow & F^N_{i \bar\imath }  , \\
\\
(n^N_i (1 - n^N_i) + \delta n_i \delta n_i)       &    \Longrightarrow & G^N_{i \bar\imath } \, ,
\end{array}
\end{eqnarray}
where $F^N$ and $G^N$ are the unknown quantities. 

Let us specify some rules to constrain them:
\renewcommand{\labelitemi}{$\diamond$}
\begin{itemize}
  \item {\bf Sum-rule:} 
  When $\overline{v}^{\rho\rho} = \overline{v}^{\kappa \kappa}$, then the last two terms in Eq.~(\ref{eq:edftotproj})
  should recombine to give $R^N_{i \bar \imath i \bar \imath} =n^N_i$. Accordingly, it seems reasonable to impose
  \begin{equation}
\label{eq:const1}
F^N_{i \bar\imath }  + G^N_{i \bar\imath }  = n^N_i\, .
\end{equation}   
  \item{\bf No-pairing limit:} Slater determinants belong to the Hilbert space spanned 
  by projected states. Consequently, one can interpret the 
  functional for particle-number projected wave functions as a generalization 
  of the SR-EDF theory expressed for Slater determinant, i.e.
    \begin{equation}
\label{eq:srlimit}
 {\cal E}_{N}  [\Psi_{N}] \Longrightarrow  {\cal E}_{SR}  [\Phi_{\rm SD}]\, ,
\end{equation}
as $\Phi_{N} \longrightarrow \Phi_{\rm SD}$. $\Phi_{\rm SD}$ denotes any Slater determinant. 
As a consequence, in this limit, we should have
\begin{equation}
\label{eq:cons2}
F^N_{i \bar\imath } \Longrightarrow n^0_i n^0_i, \hspace*{.5cm} G^N_{i \bar\imath } \Longrightarrow 0.
\end{equation} 
  \item {\bf Large $N$ limit:} In the limit of infinite particle number, the projected state and the reference 
  state should become identical (for instance $\delta n^0_i \Longrightarrow 0$).  Accordingly, we do expect
  \begin{eqnarray}
\begin{array}{ccc}
  \lim_{N \rightarrow + \infty}  F^N_{i \bar\imath }  & =& n^N_i n^N_i , \\
  \\   
  \lim_{N \rightarrow + \infty}  G^N_{i \bar\imath }  & =& n^N_i (1 - n^N_i) \,.
\end{array}
\end{eqnarray} 
\end{itemize} 
\renewcommand{\labelitemi}{$\bullet$}
These three constraints significantly reduce the freedom of choosing 
the form of the functional that can be used. The prescription~(\ref{eq:nidnini}) 
naturally fulfills all of them. 

\subsection{Can we use terms with non integer power of the density?}

When the effective two-body vertex depends explicitly on the density, then the energy
cannot be directly mapped on Eq.~(\ref{eq:edftotproj}). If the density dependence is in integer powers 
of the density, then one could eventually generalize the derivation of Eq.~(\ref{eq:edftotproj}) to three-body or even higher-order effective interactions. For all other forms of the density dependence, such as the widely used non-integer powers 
of the density, there is no
way to deduce an equivalent expression because the integration over gauge angles cannot be uniquely defined
from a 
mathematical point of view~\cite{Dob07,Dug09}. It is worth to mention that the same difficulty appears when the 
Coulomb exchange term is approximated using the Slater 
prescription. Above, we have shown that, with  a slight change in the functional 
used in the standard MR-EDF method, one obtain an EDF that can 
can be interpreted consistently within the usual functional approach where 
the projected state becomes a trial wave function to construct the ingredients 
of the functional.   

Guided by the setup of functional~(\ref{eq:edftotproj}), the most natural and 
simple way to extend the SR-EDF functional using density dependent two-body 
effective vertices with non-integer powers of the density 
is to directly replace the density entering in the effective
vertex by the density of the projected state, i.e.
\begin{equation}
\label{eq:twoext}
 \overline{v}^{\rho \rho}[\rho] \Longrightarrow  
 \overline{v}^{\rho \rho}[\rho^N] \, , ~~~~~  
 \overline{v}^{\kappa \kappa}[\rho] \Longrightarrow  
 \overline{v}^{\kappa \kappa}[\rho^N] \, .
\end{equation}   
Again, by doing this, we ensure that the functional used for the projected state 
is consistent with the one used in the no-pairing case (Eq.~(\ref{eq:srlimit})) and 
in the large-$N$ limit. 

\begin{figure}[t!]
\includegraphics[width=8.cm]{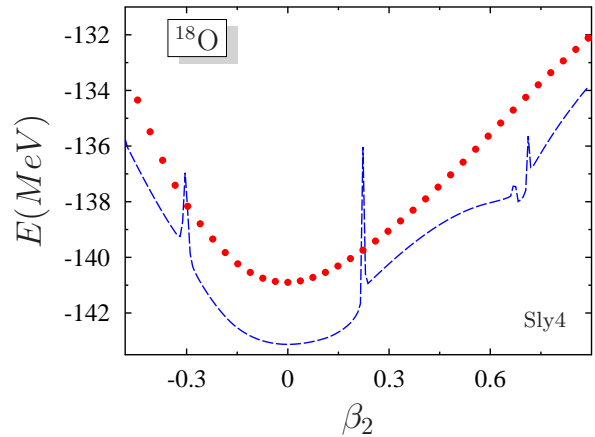}
\caption{\label{fig3:pav}
(Color online)
Same as Figure~\ref{fig1:pav} when the SLy4 effective interaction is used in the 
particle-hole channel. The dashed line corresponds to the non-regularized MR-EDF result 
directly obtained by gauge angle integration using 199 points in the discretization.
The filled circles correspond 
to the result obtained using the Symmetry-Conserved EDF using 
Eq.~(\ref{eq:edftotprojcor}) and the prescription~(\ref{eq:nidnini}). 
}
\end{figure}

In Fig.~\ref{fig3:pav}, the deformation energy curve obtained by using Eq.~(\ref{eq:twoext}) 
is  compared to the result deduced from the standard non-regularized MR-EDF
procedure  using Eq.~(\ref{eq:ekernel}). 
The SLy4 effective interaction used here contains density dependent terms with
non-integer powers i.e. $\rho^{2+1/6}$. 
Note that in this case, the MR-EDF cannot be regularized. The new alternative method we propose here, however, 
does lead  to a perfectly well behaved energy curve.

In our approach, the main difference between effective interactions that depend on non-integer
powers of the density and those depending only on integer powers of the density, is that while in the latter case one might 
eventually recognize terms coming from three-body or four-body forces and so on, this is impossible in the former case.
It should, however, be kept in mind that the use of effective interactions should be regarded 
more as a guidance for the set-up of the actual form of the functional, and not as a prerequisite for the functional approaches
as such.  

 { The example of non-integer powers of the density shows that functional theory including 
pairing and particle number restoration and extending the usual SR-EDF approach, but without using the 
MR-EDF framework,  can eventually be defined for rather general class of functionals if the strategy to 
construct the functional proposed here 
is followed. Let us add a few remarks:
\begin{itemize}
  \item It is important to realize that for particle-number projection the present strategy becomes equivalent to the MR-EDF one when the regularization
 is slightly modified compared to the one originally proposed in Ref. \cite{Lac09a}, i. e. the present strategy and the modified regularized MR EDF calculation will give the same energy for regularizable functionals. For those, it should therefore be more regarded as an alternative way of implementing MR-EDF approach to particle-number projection than as a new framework.

  \item With the present strategy, one will never have practical difficulties in applying the method
  to rather general and complex forms of functionals. However, some effort has been  made recently
  to outline the constraints that a functional should fulfill to be truly regarded as a symmetry-conserving 
  functional~\cite{Dug10}. While these constraints are even partially unknown, one might anticipate that 
  they will significantly restrict the functional form that might be used. 
   We are therefore facing the following dilemma: from condensed matter physics, we know 
  that the powerfulness of DFT comes from the large
  flexibility in choosing the functional form. 
Putting too many formal constraints will
make it increasingly difficult to model the relevant physics with a
tractable functional.  In particular, one can already see from Ref.~\cite{Dug10} that a functional that
  fulfills the constraints elaborated there will be very 
  close to the energy functional one obtains from an Hamiltonian. 
 \item It should be mentioned that the use of the projected density entering effective 
  density-dependent vertices 
 $\bar v[\rho^N]$ has already been proposed and used in Refs.~\cite{Egi91,Rod10,Rod07}. 
 However, in those references, a hybrid approach is set up where 
 transition densities are used in other parts of the functional, and for the restoration 
  of spatial symmetries.\footnote{For the 
  special case of a pure particle-number projected MR-EDF calculation, the functional used 
  in~\cite{Val97,Val00,Rod07} 
  could be mapped on a functional of the one- and two-body density matrices of the projected state.} 
  It has been pointed out in Ref.~ \cite{Rob10} that such hybrid approach may lead to unphysical 
  results when set up for the restoration of spatial symmetries. Here, the theory is completely 
  formulated in terms of the projected one- and two-body density matrices only. An open question 
  that has to be addressed in the future is if and how the strategy to set up the functional we 
  propose here can be generalized to the restoration of spatial symmetries, and perhaps even more 
  general configuration mixing without becoming numerically intractable.


\end{itemize}
}

\section{Discussion and Conclusion}

In this work, projection made by MR-EDF techniques including the recently proposed 
regularization~\cite{Lac09a,Ben09,Dug09} is further analyzed for the case of particle-number 
restoration of quasi-particle vacua of Bogoliubov type. Starting from 
a simple toy functional where the two-body effective interaction 
is not explicitly density dependent, we show that the regularized energy can {\it almost} be regarded 
as a functional of the one- and two-body densities of the projected state.
To follow the density functional methodology, we propose to slightly modify the 
functional such that it becomes a function of projected state densities {\it only},   
and that the projected state becomes the intermediate
trial state from which the functional and other
observables are constructed. For particle-number projection,
such a modification could for instance be achieved
within standard MR-EDF by slightly modifying the regularization
proposed in Ref. \cite{Lac09a} while still removing the
pathologies. Such an alternative interpretation may
eventually serve as a justification of MR-EDF framework
within a functional approach for particle number restoration
when the effective kernels are not density dependent.

As a matter of fact, most of the functional forms used nowadays do not enter into the class 
of functionals which can be safely used in MR-EDF. We show, however, that such functional 
can still be used in a symmetry restoring framework that does not make use of the MR-EDF
 technique, but directly formulates the theory in terms of the one- and two-body density matrices of projected product states.
 
This 
theory can be seen as a direct extension of the
SR-EDF level that we proposed and is called here Symmetry-Conserving EDF approach. An 
illustration of the resulting projected 
energy is given, showing that the method could be a valuable tool for the description of the ground
state of a system with pairing including the restoration of particle number even when density dependence with non-integer 
powers is used in the functional.

The analysis of similarities and differences between the MR-EDF theory and symmetry-conserving approaches was greatly simplified here
because the original quasi-particle state and the projected state share the same 
canonical basis. For instance, expression~(\ref{eq:edftotprojcor})  only holds in the canonical basis.     
In the present article, the applications are restricted to projection after variation for which this 
equation is perfectly suited.  
The next the step will be the extension approach to perform variation after projection (VAP). 
VAP is usually solved using MR-EDF techniques by making variations with respect to the components of 
the original quasi-particle vacuum and not the projected state 
itself~\cite{Egido1982,Egido1982a,She00,Sheikh2001}. 
In the symmetry conserving approach, one could follow the same strategy as in the standard 
MR-EDF approach, i.e.\ perform variations of the reference state.
Work in that direction is currently underway.

Last, we would like to mention that the present article only discusses the case of particle-number projection and 
the possibility to determine the ground-state energy. The MR-EDF technique is frequently used to restore other symmetries 
and to calculate excited states in a Generator-Coordinate framework. What these other configuration mixings have in common, is the fact that there does not exist a common canonical basis in which the one-body density matrices of 
the original and of the correlated states are simultaneously diagonal. An important point to be clarified is if and how the formalism developed here can be generalized to those more general configuration mixings, and that without becoming numerically 
intractable.  Finally, it has to be stressed that the method proposed and explored here is not meant to replace the MR EDF framework for the description of excited states and transition moments in complex nuclei. Instead, it might provide a numerically much more efficient alternative to the MR EDF scheme when one is interested just in the ground state and its evolution, either in dynamics or thermodynamics.

\begin{acknowledgments}
 We thank Thomas Duguet for stimulating discussions.
\end{acknowledgments}

\appendix

\section{Proof of Eq. (\ref{eq:edftotprojcor})}
\label{app:correctedmr}

To prove Eq.~(\ref{eq:edftotprojcor}), we have to explicitly remove terms that cause 
pathologies from the energy calculation as proposed in Ref.~\cite{Lac09a}. 
Starting from Eq.~(50) of Ref.~\cite{Lac09a}, the transition matrix elements can be expressed as
\begin{eqnarray}
\label{eq:rho01rho0}
n^{0 \varphi}_i 
& \equiv & n^0_i + \delta n_i \left[ \varphi \right], \nonumber \\
\kappa^{0 \varphi}_{i \bar\imath }
& \equiv & \kappa^0_{i \bar\imath } + \delta \kappa_{i \bar\imath } \left[ \varphi \right] , \nonumber \\
{\kappa^{\varphi 0 }_{i \bar\imath }}^\star & \equiv & \kappa^{0*}_{i \bar\imath } +  \delta \kappa_{i \bar\imath }^{ \star}\left[ \varphi \right]\, \, \, ,
\end{eqnarray}
where $n^0_i$ and $\kappa^0_{i \bar\imath }$ refer to the occupation probabilities and 
anomalous densities of the state $\Phi_0$. Following Ref.~\cite{Lac09a}, we decompose the energy kernels 
entering into the integral of Eq. ({eq:ekernel})
into three terms ${\cal E}^\rho$, ${\cal E}^{\rho \rho}$ and ${\cal E}^{\kappa \kappa}$ corresponding 
to the kinetic, mean-field and pairing terms respectively. 
Then, ${\cal E}^{\rho \rho}$ and ${\cal E}^{\kappa \kappa}$ 
can be expressed as
\begin{eqnarray}
{\cal E}^{\rho \rho} & = & \frac{1}{2} \sum_{ij} \bar{v}_{ijij}^{\rho \rho} n^0_i n^0_j \nonumber \\
 & +& \frac{1}{2} \sum_{ij} \bar{v}_{ijij}^{\rho \rho} \left(n^0_i  \delta n_j \left[ \varphi \right]  + n^0_j \delta n_i \left[ \varphi \right] \right)\nonumber \\
 &+&  \frac{1}{2} \sum_{ij} \bar{v}_{ijij}^{\rho \rho} \delta n_i \left[ \varphi \right] \delta n_j \left[ \varphi \right] \, ,
 \label{eq:drhodrho}
\end{eqnarray} 
whereas
\begin{eqnarray}
{\cal E}^{\kappa \kappa} & = &
\frac{1}{4} \sum_{ij} \bar{v}_{i{\bar\imath }j{\bar\jmath }}^{\kappa \kappa} \, \kappa^{0 *}_{i{\bar\imath }} \, \kappa^{0}_{j{\bar\jmath }} \nonumber \\
&+& \frac{1}{4} \sum_{ij} \bar{v}_{i{\bar\imath }j{\bar\jmath }}^{\kappa \kappa} \, 
\left( \kappa^{0 *}_{i{\bar\imath }} \, \delta \kappa_{j \bar\jmath } \left[ \varphi \right] + \kappa^{0}_{j{\bar\jmath }}  \delta \kappa^{\star}_{i \bar\imath } \left[ \varphi \right] \right) \nonumber \\
&+& \frac{1}{4} \sum_{ij} \bar{v}_{i{\bar\imath }j{\bar\jmath }}^{\kappa \kappa} \, 
 \delta \kappa^{\star}_{i \bar\imath } \left[ \varphi \right] \delta \kappa_{j \bar\jmath } \left[ \varphi \right] \, .
 \label{eq:dkappadkappa}
\end{eqnarray} 
These expressions are the strict equivalent 
of the ones given in Eqns.~(51-54) in Ref.~\cite{Lac09a}. 
For instance, regularizations have been obtained by removing terms with $j = \bar\imath $ in the 
last line of Eqs. (\ref{eq:drhodrho}) and Eq. (\ref{eq:dkappadkappa}).
Accordingly, the spurious contribution to be removed from the functional is
\begin{eqnarray}
\label{eq:cor_rr} {\cal E}^{\rho \rho}_{CG} & = & \frac{1}{2}  \sum_{i} \bar{v}_{ijij}^{\rho \rho} \,
\int \delta n_{i}\left[ \varphi \right] \delta n_{i}\left[ \varphi \right] {\cal N}_N({0,\varphi})d\varphi 
       \, ,\nonumber \\
       \\
\label{eq:cor_kk} {\cal E}^{\kappa \kappa} _{CG}& = & \frac{1}{2} \sum_{i}
\bar{v}_{i{\bar\imath }i{\bar\imath }}^{\kappa \kappa}  \,\int 
 \delta \kappa^{\star}_{i \bar\imath } \left[ \varphi \right] \delta \kappa_{i \bar\imath } \left[ \varphi \right] 
 {\cal N}_N({0,\varphi})d\varphi \, .\nonumber \\
\end{eqnarray}
Therefore, when the regularization is included, this is equivalent to make the replacements
 \begin{eqnarray}
&&  \int_{0}^{2\pi} \!\!\! d\varphi  ~n^{0\varphi}_i ~n^{0\varphi}_i   {\cal N}_N({0,\varphi}) \nonumber \\
&& \Longrightarrow  \int_{0}^{2\pi} \!\!\! d\varphi  ~\left( 
n^{0\varphi}_i ~n^{0\varphi}_i  - \delta n_{i}\left[ \varphi \right] \delta n_{i}\left[ \varphi \right] \right) {\cal N}_N({0,\varphi})  
\nonumber 
\end{eqnarray}
and 
\begin{eqnarray}
&& \int_{0}^{2\pi} \!\!\! d\varphi {\kappa^{\varphi 0}_{i \bar\imath }}^\star {\kappa^{0\varphi}_{i \bar\imath }}  {\cal N}_N({0,\varphi})  \nonumber \\
&& \Longrightarrow  \int_{0}^{2\pi} \!\!\! d\varphi \left( 
{\kappa^{\varphi 0}_{i \bar\imath }}^\star {\kappa^{0\varphi}_{i \bar\imath }} 
-  \delta \kappa^{\star}_{i \bar\imath } \left[ \varphi \right] \delta \kappa_{i \bar\imath } \left[ \varphi \right] \right)  {\cal N}_N({0,\varphi}) \, , \nonumber 
\end{eqnarray}
in the last two terms of Eq.~(\ref{eq:edftotproj}).

From the equalities (\ref{eq:rho01rho0}), one can deduce new interesting 
relationships between the projected observables. For instance, performing the gauge integration of 
the first equation, we obtain 
\begin{equation}
n^N_i = \int_{0}^{2\pi} \!\!\! d\varphi  ~n^{0\varphi}_i   {\cal N}_N({0,\varphi})  =n^0_i + \delta n_i \, ,
\end{equation}
with 
\begin{equation}
\delta n_i  = n^N_i -n^0_i = \int d\varphi \, \delta n_{i}\left[ \varphi \right]  {\cal N}_N({0,\varphi}) \, .
\end{equation}
From this, let us now re-express the different quantities entering in Eq.~(\ref{eq:edftotproj})
\begin{eqnarray}
\int_{0}^{2\pi} \!\!\! d\varphi  ~n^{0\varphi}_i ~n^{0\varphi}_j   {\cal N}_N({0,\varphi}) & = & n^0_i n^0_j \nonumber \\
&+& n^0_i \delta n_j +  \delta n_i  n^0_j \nonumber \\
&+& 
\int_{0}^{2\pi} \!\!\! d\varphi  ~\delta n_i\left[ \varphi \right]  ~ \delta n_j \left[ \varphi \right]   {\cal N}_N({0,\varphi}) \, ,\nonumber 
\end{eqnarray}
where, in the specific case $i=j$, we recognize the term that enters in the regularization to 
be the last one. Therefore, the term entering into the regularization of  ${\cal E}^{\rho\rho}$ 
can be expressed as
\begin{eqnarray}
&& \int_{0}^{2\pi} \!\!\! d\varphi  ~\left( 
n^{0\varphi}_i ~n^{0\varphi}_i  - \delta n_{i}\left[ \varphi \right] \delta n_{i}\left[ \varphi \right] \right) {\cal N}_N({0,\varphi}) 
 \nonumber \\
&& \hspace*{3.cm} = n^0_i n^0_i + 2 n^0_i \delta n_i \nonumber \\
&& \hspace*{3.cm} =n^N_i n^N_i - \delta n_i \delta n_i \, .\label{eq:funnini0}
\end{eqnarray}

To derive an expression of the term entering in the regularization of ${\cal E}^{\kappa\kappa}$, one can proceed 
in a similar way.
We first define $\delta \kappa_{i \bar\imath }^{ *}$ and $\delta \kappa_{i \bar\imath }$ through 
\begin{eqnarray}
\int_{0}^{2\pi} \!\!\! d\varphi  ~ {\kappa^{\varphi 0 }_{i \bar\imath }}^\star {\cal N}_N({0,\varphi})& = & 
\int_{0}^{2\pi} \!\!\! d\varphi  \left(\kappa^{0 *}_{i \bar\imath } + \delta \kappa_{i \bar\imath }^{\star}\left[ \varphi \right] \right) {\cal N}_N({0,\varphi})\nonumber  \\
&\equiv&  \kappa^{0 *}_{i \bar\imath } + \delta \kappa_{i \bar\imath }^{\star} \nonumber  \\
\int_{0}^{2\pi} \!\!\! d\varphi  \kappa^{0 \varphi}_{i \bar\imath }{\cal N}_N({0,\varphi})
& = & \int_{0}^{2\pi} \!\!\! d\varphi  \left(\kappa^0_{i \bar\imath } + \delta \kappa_{i \bar\imath } \left[ \varphi \right] \right) {\cal N}_N({0,\varphi})\nonumber  \\
&\equiv& \kappa^0_{i \bar\imath } + \delta \kappa_{i \bar\imath } \, .  \nonumber  
\end{eqnarray}
Therefore the term entering in the regularized functional is given by
\begin{eqnarray}
&& \int_{0}^{2\pi} \!\!\! d\varphi  ~\left( 
{\kappa^{\varphi 0}_{i \bar\imath }}^\star {\kappa^{0\varphi}_{i \bar\imath }}  
-\delta \kappa_{i \bar\imath }^{ \star} \left[ \varphi \right]  \delta \kappa_{i \bar\imath } \left[ \varphi \right]  
\right) {\cal N}_N({0,\varphi}) 
 \nonumber \\
&& \hspace*{1.cm} = \kappa^{0 *}_{i \bar\imath } \kappa^{0}_{i \bar\imath } + \delta \kappa^{\star}_{i \bar\imath }  \kappa^0_{i \bar\imath } + \kappa^{0 *}_{i \bar\imath } \delta \kappa_{i \bar\imath } \, .
\end{eqnarray}
One can then take advantage of the fact that 
\begin{eqnarray}
n^N_i &=& n^0_i n^0_i + 2  n^0_i \delta n_i \nonumber \\
&+& 
\int_{0}^{2\pi} \!\!\! d\varphi  ~\delta n_i\left[ \varphi \right]  ~ \delta n_j \left[ \varphi \right]   {\cal N}_N({0,\varphi}) \nonumber \\
&
+ & \kappa^{0 *}_{i \bar\imath } \kappa^{0}_{i \bar\imath } + \delta \kappa^{\star}_{i \bar\imath }  \kappa^0_{i \bar\imath } + \kappa^{0 *}_{i \bar\imath } \delta \kappa_{i \bar\imath }
\nonumber \\
&+& \int_{0}^{2\pi} \!\!\! d\varphi   
\delta \kappa_{i \bar\imath }^{ \star} \left[ \varphi \right]  \delta \kappa_{i \bar\imath } \left[ \varphi \right]  {\cal N}_N({0,\varphi}) \nonumber 
\end{eqnarray}
and that
\begin{eqnarray}
\delta n_i\left[ \varphi \right]\delta n_i\left[ \varphi \right] = -  \delta \kappa_{i \bar\imath } \left[ \varphi \right] 
\delta \kappa_{i \bar\imath }^{ \star} \left[ \varphi \right].
\end{eqnarray}
The first equality is nothing but Eq.~(\ref{eq:sumrule}), whereas the second equality can be proved by expressing
$\delta n_i\left[ \varphi \right]$,  $\delta \kappa_{i \bar\imath } \left[ \varphi \right] $ and 
$\delta \kappa_{i \bar\imath }^{ *} \left[ \varphi \right]$ directly in terms of the $u_i$ and 
$v_i$ of the SR-EDF theory  and the gauge angle $\varphi$ starting from Eq.~(72-74)
of~Ref.~\cite{Lac09a}. Altogether, we obtain:
\begin{eqnarray}
&& \int_{0}^{2\pi} \!\!\! d\varphi  ~\left( 
{\kappa^{\varphi 0}_{i \bar\imath }}^\star {\kappa^{0\varphi}_{i \bar\imath }}  
-\delta \kappa_{i \bar\imath }^{ \star} \left[ \varphi \right]  \delta \kappa_{i \bar\imath } \left[ \varphi \right]  
\right) {\cal N}_N({0,\varphi}) = n^N_i 
 \nonumber \\
&& -\int_{0}^{2\pi} \!\!\! d\varphi  ~\left( 
n^{0\varphi}_i ~n^{0\varphi}_i  - \delta n_{i}\left[ \varphi \right] \delta n_{i}\left[ \varphi \right] \right) {\cal N}_N({0,\varphi})  \nonumber \\
&& = (n^N_i (1 - n^N_i) + \delta n_i \delta n_i)\, . \label{eq:corkk}
\end{eqnarray}  
Combining this expression with Eq.~(\ref{eq:funnini0}), we finally deduce the 
expression~(\ref{eq:edftotprojcor}) for the regularized functional.

\section{Particle number restoration with recurrence relation}
\label{app:rec}

A method, alternative to the the gauge-integration method is presented here to
calculate the one- and two-body density matrix components of a projected product state. 
This method turns out to be very fast and efficient numerically.

Let us start from a quasi-particle state written in its canonical basis as
\begin{eqnarray}
| \Phi_{0} \rangle  & = & \prod_{i>0} \left( 1 + x_i a^\dagger_i a^\dagger_{\bar\imath } \right) | 0 \rangle \, ,
\label{eq:bcsstate}
\end{eqnarray}
where $|x_i|^2 = n_i^0/ (1-n_i^0)$. The associated projected state with $N$ particles can be expressed as
\begin{eqnarray}
| \Psi_N \rangle  &  \propto & \left( \sum_{i>0} x_i a^\dagger_i a^\dagger_{\bar\imath } \right)^N | 0 \rangle \, .
\label{eq:pbcsstate}
\end{eqnarray}  
Starting from these expressions, it has recently been shown \cite{Lac10}
that the elements of the one- and two-body density matrix 
are given by
\begin{eqnarray}
n^N_i &=& \displaystyle  N |x_i|^2 \frac{I_{N-1}(i)}{I_N} \, ,   \label{eq:nin}\\
R^N_{i\bar\imath  j \bar\jmath } &=&  \displaystyle N x^*_i x_j \frac{I_{N-1}(i,j)}{I_N} ~~{\rm for} ~~ (i \neq j) \, ,  \\
R^N_{ijij} &=&  N(N-1) |x_i|^2 |x_j|^2 \frac{I_{N-2}(i,j)}{I_N}  \, ,
\label{eq:nicipbcs}
\end{eqnarray} 
while as already mentioned $R^N_{i\bar\imath  i \bar\imath } = n^N_i$. The different coefficients 
entering in $n^N$ and $R^N$ are given by:
\begin{eqnarray}
\left\{
\begin{array}{ccc}
I_K &=&   \sum^{\neq}_{(i_1, \cdots ,i_{K})} |x_{i_1}|^2 \cdots |x_{i_{K}}|^2 \\
\\
I_K(i) &=& \sum^{\neq}_{(i_1, \cdots ,i_{K}) \neq i} |x_{i_1}|^2 \cdots |x_{i_{K}}|^2  \\
\\
I_K(i,j) &=& \sum^{\neq}_{(i_1, \cdots ,i_{K}) \neq (i,j)} |x_{i_1}|^2 \cdots |x_{i_{K}}|^2\nonumber \\
\nonumber \\
& \cdots & \nonumber
\label{eq:ikijdef}
\end{array}
\right. 
\end{eqnarray} 
Direct use of these expressions for $K=N$ is rather difficult numerically. However, these
coefficients verify simple recurrence relations that are straightforward to implement on a computer.
These recurrence relations have been recently used to solve numerically the Variation After Projection (VAP)
 \cite{San08,San09} and to set up a new functional 
for pairing accounting for particle-number conservation~\cite{Lac10}. 

In the present work, we  use the recurrence method to perform PAV within the symmetry-conserving 
EDF framework. In that case, a preliminary SR-EDF calculation is performed leading to a 
quasi-particle state given by~(\ref{eq:bcsstate})
with a set of $\{ x_i \}$ values. Here, we have used the ev8 code~\cite{Bon05}. 
From this set, the   
quantities $I_{N-1}(i)$ and $I_{N}$ are evaluated via the recurrence relations
\begin{eqnarray}
I_{K}(i) &=& I_{K} - (K-1) |x_i|^2 I_{K-1}(i) \nonumber\\
I_{K} &=& \sum_i |x_i|^2 I_{K-1} - (K-2)  \sum_i |x_i|^4 I_{K-2}(i) \, . \nonumber\\
\end{eqnarray}
with the condition $I_0 = I_0 (i) = 1 $, $I_1 = \sum_k |x_k|^2 $ and $I_1(i) = I_1 - |x_i|^2$. 
The occupation numbers of the projected state can then be calculated as well as the correlation 
components using the relation~\cite{Lac10,Hup10}:  
\begin{eqnarray}
R^N_{i\bar\imath  j \bar\jmath } &=&  \displaystyle  x^*_i x_j \frac{n_j^N - n_i^N}{|x_j|^2 - |x_i|^2} ~~{\rm for} ~~ (i \neq j) \, ,  \\
R^N_{ijij} &=&  \frac{|x_j|^2 n_i^N - |x_i|^2 n_j^N}{|x_j|^2 - |x_i|^2}  \, ,
\label{eq:pratikpbcs}
\end{eqnarray}
where for $i=j$, we have $R^N_{i\bar\imath  j \bar\jmath } = n_i^N$ and $R^N_{ijij} = 0$. This method 
is referred to as "recurrence method" in the text.

\bibliographystyle{apsrev4-1}
\bibliography{PAV-func}

\end{document}